\documentclass{aa}  
\usepackage{graphicx}
\usepackage{txfonts}
\usepackage{lipsum}
\usepackage{subcaption}
\usepackage{xcolor}
\usepackage{lscape}
\usepackage{placeins}

\usepackage{natbib}
\bibpunct{(}{)}{;}{a}{}{,}
\usepackage[breaklinks, colorlinks, citecolor=blue, linkcolor=blue]{hyperref}
\usepackage{orcidlink}

\begin{document}
    \title{Atmospheric constraints on GJ\,1214\,b from CRIRES$^+$ and prospects for characterisation with ANDES}
    \titlerunning{Atmospheric constraints on GJ\,1214\,b}\authorrunning{A. Peláez-Torres et al.}

   \author{A.\,Peláez-Torres
          \inst{\ref{IAA_CSIC}}\thanks{\texttt{E-mail: \color{magenta}apelaez@iaa.es}}\,\orcidlink{0000-0001-9204-8498},
          A.\,Sánchez-López\inst{\ref{IAA_CSIC}}\thanks{\texttt{E-mail: \color{magenta}alexsl@iaa.es}}\,\orcidlink{0000-0002-0516-7956},
          C.\,Jiang\inst{\ref{IAC},\ref{ULL}},
          E.\,Pallé\inst{\ref{IAC},\ref{ULL}},
          J.\,Orell-Miquel\inst{\ref{Austin}},
          M.\,López-Puertas\inst{\ref{IAA_CSIC}},  
          L.\,T.\,Parker\inst{\ref{Oxford}},  
          and H.\,Diamond-Lowe\inst{\ref{Denmark},\ref{Baltimore}}
          }

    \institute{Instituto de Astrofísica de Andalucía (IAA-CSIC), Glorieta de la Astronomía s/n, Genil, E-18008 Granada, Spain \label{IAA_CSIC}
    \and Instituto de Astrof\'isica de Canarias (IAC), 38200 La Laguna, Tenerife, Spain \label{IAC}
    \and Departamento de Astrof\'isica, Universidad de La Laguna (ULL), 38206 La Laguna, Tenerife, Spain \label{ULL}
    \and Department of Astronomy, University of Texas at Austin, 2515 Speedway, Austin, TX 78712, USA \label{Austin}
    \and Department of Astrophysics, University of Oxford, Denys Wilkinson Building, Keble Road, Oxford OX1 3RH, UK \label{Oxford}
    \and Department of Space Research and Space Technology, Technical University of Denmark, Elektrovej 328, DK-2800 Kgs. Lyngby, Denmark \label{Denmark}
    \and Space Telescope Science Institute, 3700 San Martin Drive, Baltimore, MD 21218, USA \label{Baltimore}
             }

   \date{Received 05 December 2025}

 \abstract
{Sub-Neptune exoplanets like GJ\,1214\,b provide a critical link between terrestrial and giant planets, yet atmospheric characterisation remains challenging due to high-altitude clouds and compressed atmospheres. JWST has recently hinted at molecular signals in GJ\,1214\,b, and ground-based high-resolution spectroscopy is potentially able to confirm them.}
{We aim to constrain the atmospheric composition of GJ\,1214\,b using all available transits observed with the upgraded CRIRES$^+$ spectrograph at the Very Large Telescope (VLT) by searching for the signatures of water vapour, methane, and carbon dioxide.} 
{We analysed eight CRIRES$^+$ transit datasets covering the K band ($1.90$\,--\,$2.45$\,$\mu$m) at a resolving power of $\mathcal{R}$\,$\approx$\,$100{,}000$. We used the SysRem algorithm to correct for telluric and stellar contributions and employed the cross-correlation technique with templates from petitRADTRANS to search for H$_2$O, CH$_4$, and CO$_2$. Injection–recovery tests across a grid of metallicities ($Z$) and cloud-deck pressures ($p_c$) were performed to quantify detection limits. We also generated predictions for ANDES observations using end-to-end simulated datasets with {\tt EXoPLORE}.}
{We detect no significant H$_2$O, CH$_4$, or CO$_2$ signatures. Injection–recovery tests show that such non-detections exclude atmospheres with low-altitude clouds and moderate or low metallicities. CH$_4$ yields the tightest empirical limits, with CO$_2$ unexpectedly ruling out intermediate metallicities ($\sim$\,$100\times$\,solar) with clouds deeper due to its rapidly rising opacity in compressed, high-$Z$ atmospheres. Our constraints are in line with either a high-$Z$ or a high-altitude aerosol layer, in agreement with recent JWST inferences.}
{The combined analysis of eight CRIRES$^+$ datasets provides the most stringent high-resolution constraints on the atmospheric properties of GJ\,1214\,b to date. Planetary signals are likely buried below our current detection threshold, preventing confirmation of the recent JWST-reported molecular hints. Simulations of a single transit observed with ANDES on the ELT predict modest improvements for H$_2$O, a substantially expanded detectable region for CH$_4$, and the strongest gains for CO$_2$, making the latter a particularly effective tracer for characterising high-metallicity atmospheres in sub-Neptunes.
    }


   \maketitle
   \nolinenumbers

\section{Introduction}
\label{sec:introduction}

Sub-Neptune–size planets are now recognised as one of the most common outcomes of planet formation, dominating the radius distribution between roughly $1.5$ and $4\,R_{\oplus}$ \citep{2017AJ....154..109F, 2018AJ....156..264F}. Despite their relative similar sizes, these planets seem to span a wide diversity of interior structures: some are consistent with high-density, rocky or water-rich compositions, whereas others appear to host low-density H/He envelopes \citep[e.g.,][]{2022Sci...377.1211L, parc2024super, venturini2024fading, lichtenberg2025constraining}. In principle, sub-Neptunes with H/He envelopes should imprint prominent spectroscopic features during the exoplanet's primary transit. However, a decade of observations has repeatedly yielded featureless or strongly muted transmission spectra for many of these worlds with space telescopes at low-resolution \citep[e.g.][]{kreidberg2014clouds, libby2022featureless,lustig2023jwst, kahle2025space, bennett2025hst, bennett2025additional} and ground-based, high-resolution instrumentation \citep[e.g.][]{spyratos2021transmission, jiang2023featureless, 2024MNRAS.530.3100D, 2024A&A...688A.191G, 2025MNRAS.538.3263P}. GJ\,1214\,b \citep{charbonneau2009super}, a benchmark sub-Neptune subjected to intensive atmospheric scrutiny \citep[e.g.,][]{bean2010ground, miller2010nature, kempton2011atmospheric, morley2013quantitatively, kreidberg2014clouds, 2022A&A...659A..55O, 2023Natur.620...67K, nixon2024new, 2024ApJ...974L..33S, 2025ApJ...979L...7O}, is a prime example of this phenomenon, after its featureless spectrum became emblematic of the difficulties faced in characterising this planetary population \citep{bean2010ground, kreidberg2014clouds}.

Multiple mechanisms have been proposed to explain these muted spectra. Extinction by high-altitude aerosols can obscure molecular features  \citep{morley2013quantitatively, kreidberg2014clouds}, while extremely metal-enriched atmospheres would raise the mean molecular weight high enough to suppress their amplitude \citep[e.g.,][]{2023Natur.620...67K}. Due to the frequency of featureless spectra found, both processes could be widespread across the sub-Neptune population \citep{lee2025mineral, welbanks2025challenges}. The recent arrival of JWST has dramatically improved this landscape as molecular detections (including water vapour, carbon dioxide or methane) have been confirmed in several sub-Neptunes \citep{2024arXiv240303325B, holmberg2024possible, schmidt2025comprehensive}, enabling robust constraints on atmospheric metallicities and cloud properties. However, a significant part of the observations
are still limited by apparent aerosol extinction, as it is the case for GJ 1214 b \citep{2023Natur.620...67K, 2024ApJ...974L..33S, 2025ApJ...979L...7O}, L\,98-59\,c \citep{scarsdale2024jwst}, or GJ\,3090\,b \citep{ahrer2025escaping}.

Observations using high-resolution spectroscopy (HRS, $\mathcal{R}$\,$\gtrsim$\,$40,000$) emerged as potentially one of the most effective approaches for characterising aerosol-rich exoplanet atmospheres \citep{2020AJ....160..198H}. Applying the well-known cross-correlation (CC) technique to these data, we are in principle capable of probing thin atmospheric layers above the high clouds of sub-Neptunes \citep{2018arXiv180604617B, 2025ARA&A..63...83S}. This is because HRS data allow us to resolve the core and wings of spectral lines, contrasting to low-resolution studies. This makes it an ideal technique to attempt detections in hazy exo-atmospheres, where only line-cores may protrude above the continua provided by a cloud-deck \citep[e.g.,][]{2018A&A...619A...3P, 2018A&A...612A..53P, 2020A&A...643A..24S, gao2021aerosols}. Indeed, HRS characterisation of exoplanet atmospheres has allowed the detection of single species \citep{2002ApJ...568..377C, 2008A&A...487..357S, 2010Natur.465.1049S, 2012Natur.486..502B, 2013MNRAS.436L..35B, 2013A&A...554A..82D, 2014ApJ...783L..29L, 2021A&A...654A.163C, 2022A&A...662A.101S}, and multiple molecules in the past \citep{2018ApJ...863L..11H, 2019MNRAS.482.4422C, 2020AJ....160..228K, 2021Natur.592..205G, 2021A&A...651A..33C, 2022A&A...661A..78S} at layers that can typically span pressures below $1$\,mbar (higher altitudes than this pressure level).

A particularly valuable tracer of atmospheric escape in close-in, low-density exoplanets is the metastable helium triplet at $1083$ nm, observable with high- and low-resolution spectroscopy \citep{Seager2010,oklopvcic2018new}. This feature probes the structure of the thermosphere and has revealed extended, escaping atmospheres in several sub-Neptunes and warm Neptunes \citep{spake2018helium, allart2018spectrally, nortmann2018ground}. GJ\,1214\,b \citep{charbonneau2009super} is a benchmark warm sub-Neptune ($2.7$ R$_\oplus$, $8.1$ M$_\oplus$) orbiting an M4.5 dwarf every $1.58$\,days and has been the focus of extensive observational campaigns aimed at constraining its atmospheric composition. Several high-resolution searches for He I at $1083$ nm have been conducted and can only place upper limits \citep[e.g.][]{crossfield2019upper, petit2020upper, kasper2020nondetection, 2022A&A...659A..55O, spake2022non} or a tentative detection \citep{2022A&A...659A..55O} on the planetary absorption. None have yielded a robust conclusive detection.

GJ\,1214 is not considered a highly active M dwarf in terms of strong flares or chromospheric emission, but long-term photometric monitoring and transit analyses have revealed the presence of spots and active regions on its surface \citep{berta2011gj1214, carter2011transit, narita2013multi, nascimbeni2015large, rackham2017access, mallonn2018gj}. Photospheric heterogeneity can influence transit depths and potentially introduce epoch-to-epoch variations in transmission spectroscopy. Nevertheless, there is currently no evidence that such moderate activity levels significantly modulate the He I $1083$ nm signal. 

A powerful instrument for HRS in transmission is the upgraded CRyogenic high-resolution InfraRed Echelle Spectrograph \citep[CRIRES$^+$;][]{2004SPIE.5492.1218K, 2014SPIE.9147E..19F, 2014Msngr.156....7D}. Thanks to its high spectral resolution of $\mathcal{R}$\,$\approx$\,$85\,000$--$100\,000$, CRIRES$^+$ is in principle sensitive enough to detect the strong spectral features that form at low pressures above potential clouds, depending on the actual conditions in the exoplanet. In this paper, we compiled all available datasets observed with CRIRES$^+$ for GJ\,1214\,b (eight transits in total) to explore its atmospheric composition. In Sect.\,\ref{sec:observations}, we describe these observations. The methods used for the analysis of the spectral data are based on the well-known cross-correlation technique, which is detailed in Sect.\,\ref{sec:methods}. We present and discuss our results in Sect.\,\ref{sec:results}, including prospects for atmospheric characterisation of GJ\,1214\,b with the next generation of telescopes and instruments like ANDES \citep{marconi2022andes, 2025ExA....59...29P}, and we outline our main conclusions in Sect.\,\ref{sec:conclusions}.

\section{Observations}
\label{sec:observations} 

\begin{figure}
    \centering
    \includegraphics[width=\columnwidth]{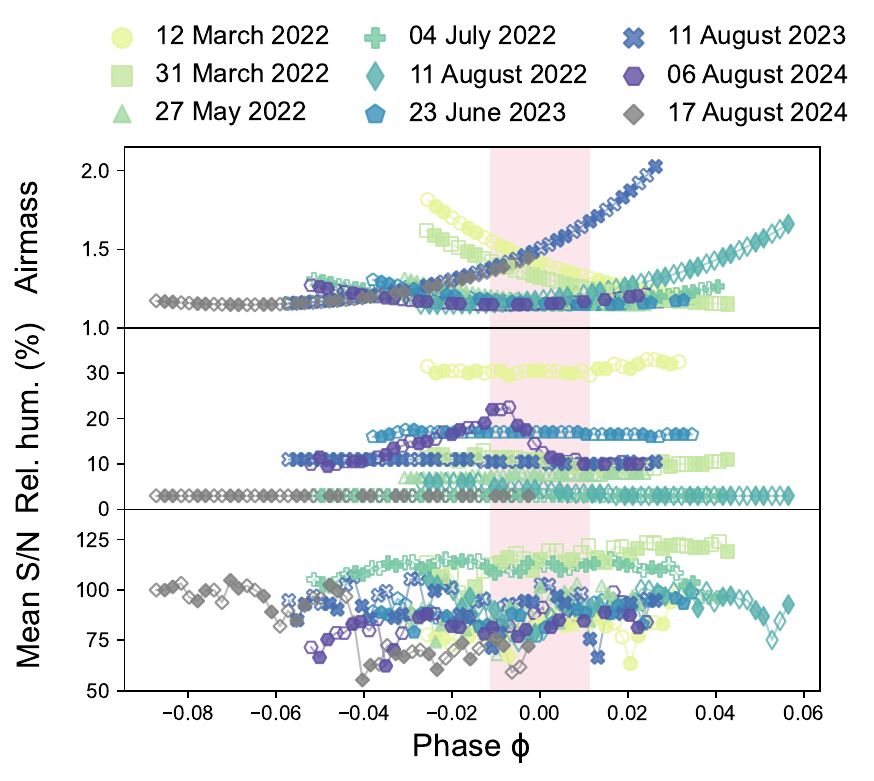}
    \caption{Evolution of the airmass, relative humidity, and mean S/N ratio per spectrum. The pink shaded area marks the in-transit times. Nodding A spectra are represented with empty symbols and Nodding B spectra with filled symbols.}
    \label{fig:night_conditions}
\end{figure}

We analysed eight datasets from the public ESO archive observed with the CRIRES$^+$ spectrograph, installed at the 8.2-m Unit Telescope 3 of the ESO's Very Large Telescope (VLT) at Cerro Paranal. The observations were carried out between March $12$ $2022$ and August $06$ $2024$ in the K band, using the K2148 wavelength setting. This setting covers the $1.90$\,--\,$2.45$\,$\mu$m range over eight spectral orders, at a resolving power of $\mathcal{R}$\,$\approx$\,$100\,000$. In Table\,\ref{tab:planets_params} we summarised the stellar and planetary parameters used in this study. The eight analysed datasets covers the pre-, during-, and post-transit phases of GJ\,1214\,b, from $\phi=-0.0114$ to $\phi=0.0114$; (see Fig.\,\ref{fig:night_conditions}). A ninth night (August $17$, $2017$) was available, but its low S/N and poor coverage of the transit event made it unsuitable for inclusion in our analysis. The S/N across all eight nights ranges from $\sim$\,$55$ to $\sim$\,$120$. All datasets show relative humidity below $20$\%, except for the night of March $12$, $2022$, when it was about $30$\%. Further details on the observing conditions of this night are provided in Table\,\ref{tab:observations}. 

The CRIRES$^+$ raw data for GJ\,1214\,b were retrieved from the ESO Archive\footnote{\url{https://archive.eso.org/cms/eso-data.html}} and reduced using the standard \texttt{CR2RES} pipeline routines (version $1.4.4$) through the ESO Recipe Execution Tool (\texttt{EsoRex}, version $3.13.8$). This pipeline includes dark calibration (not used during nodding subtraction), bad-pixel masking, flat calibration, wavelength calibration using a combination of the Fabry-Pérot etalon and uranium-neon lamp, and 1D spectral extraction.
The observations consisted of $7.5$ nodding cycles, following the ``$\rm A_{1,2,3}$,\,$\rm B_{1,2,3}$,\,$\rm B_{4,5,6}$,\,$\rm A_{4,5,6}$'' nodding pattern. In order to perform nodding subtraction, we regrouped the nodding pairs as $\rm A_1B_1$, $\rm A_2B_2$, ... $\rm A_6B_6$. The spectra were extracted using the optimal extraction algorithm, with a fixed extraction height of $20$ pixels ($\sim8$ times the median full width at half maximum of the point spread function along the slit) and no-light rows subtracted. Also, we discarded the bluest order (order $29$; $1.90$--$1.95$ $\mu$m) due to strong telluric absorption. The observations were conducted in adaptive-optics mode and, due to seeing conditions ($\sim$\,$0.65''$), the slit was not evenly illuminated in the dispersion direction during the observations. Although this might cause a slight spectral shift between A and B positions (as described in the \texttt{CR2RES} user manual\footnote{\url{https://ftp.eso.org/pub/dfs/pipelines/instruments/cr2res/cr2re-pipeline-manual-1.6.10.pdf}}), no significant spectral shift above the velocity resolution of the instrument ($1$\,km\,s$^{-1}$) was detected when cross-correlating the A and B spectra.

\begin{table}
\renewcommand{\arraystretch}{1.3}
\centering
\caption{Stellar and planet parameters of the GJ\,1214 system.}
\label{tab:planets_params}
\begin{tabular}{l c c}
\hline
\hline
\noalign{\smallskip}
Parameter & Value & Reference \\
\noalign{\smallskip}
\hline
\noalign{\smallskip}

\noalign{\smallskip}
\multicolumn{2}{c}{\textit{Stellar parameters} } \\
\noalign{\smallskip}
$M_{\star}\,(M_{\odot})$ & $0.1820^{+0.0042}_{-0.0041}$ & M24 \\
$R_{\star}\,(R_{\odot})$ & $0.2162^{+0.0025}_{-0.0024}$ & M24 \\
$T_{\rm eff}$\,(K) & $3101\pm43$ & M24 \\
$d$\,(pc) & $14.642\pm0.018$ & M24 \\
\noalign{\smallskip}
\multicolumn{2}{c}{\textit{Planet parameters} } \\
\noalign{\smallskip}
$M_{\rm p}$\,(M$_{ \oplus}$) & $8.41^{+0.36}_{-0.35}$ & M24 \\
$R_{\rm p}\,(R_{\oplus})$ & $2.733^{+0.033}_{-0.031}$ & M24 \\
$\rho_{\rm p}$\,(g $\rm cm^{-3}$) & $2.26\pm0.11$ & M24 \\
$K_{\rm p}$\,(km\,s$^{-1}$) & $106.68\pm6.78$ & This work \\
$T_{\rm eq}$\,(K) & $567.0\pm7.6$ & M24 \\
\noalign{\smallskip}
\multicolumn{3}{c}{\textit{Transit and system parameters} } \\
\noalign{\smallskip}
$P$\,(d) & $1.580404531^{+0.000000018}_{+0.000000017}$ & M24 \\
$t_0$\,(BJD)& $2459782.0176719\pm0.0000083$ & K23 \\
$t_{\rm 14}$\,(days) & $0.036236^{+0.000046}_{-0.000045}$ & M24 \\
$a$\,(au) & $0.01505\pm0.00011$ & M24 \\
$i$\,(deg) & $88.980^{+0.094}_{-0.085}$ & M24 \\
$e$ & $0.0062^{+0.0079}_{-0.0044}$ & M24 \\
$K$\,(m\,s$^{-1}$) & $14.38^{+0.57}_{-0.56}$ & M24 \\
\noalign{\smallskip}
\hline
\end{tabular}
\tablefoot{M24: \cite{2024ApJ...963L..37M}, and K23: \cite{2023Natur.620...67K}.}
\end{table}

\begin{table*}
\caption[]{GJ\,1214's CRIRES$^+$ analysed spectroscopic observations.}
\label{tab:observations}
\centering
\begin{tabular}{lllllllc}
  \hline \hline
  \noalign{\smallskip}
  UT date & N$_{\rm exp}$ & T$_{\rm exp}$\,(s) & Airmass & Rel. hum\,(\%) & Mean S/N & Discarded orders & Discarded spectra \\
  \noalign{\smallskip}
  \hline
  \noalign{\smallskip}
  $2022$ March $12$ & $32$ & $240$ & $1.43$ & $31$ & $81$ & $29$ & $5$, $9$, $19$, $21$, $27$ \\
  $2022$ March $31$ & $38$ & $240$ & $1.31$ & $10$ & $116$ & $29, 28$ & $3$, $8$, $10$, $29$ \\
  $2022$ May $27$ & $40$ & $180$ & $1.20$ & $8$ & $83$ & $29$ & - \\
  $2022$ July $04$ & $50$ & $240$ & $1.20$ & $3$ & $110$ & $29$ & - \\
  $2022$ August $11$ & $46$ & $240$ & $1.29$ & $4$ & $93$ & $29$ & $2$, $5$ \\
  $2023$ June $23$ & $40$ & $240$ & $1.18$ & $17$ & $91$ & $29$ & $3$, $4$, $12$, $13$, $28$, $32$, $33$, $36$ \\
  $2023$ August $11$ & $46$ & $240$ & $1.42$ & $11$ & $93$ & $29$ & $1$, $20$ \\
  $2024$ August $06$ & $40$ & $240$ & $1.18$ & $14$ & $83$ & $29$ & $9$, $10$ \\
  \noalign{\smallskip}
  \hline
\end{tabular}
\tablefoot{PI of the nights of March $31$ $2022$ (108.22PH.005), July $04$ $2022$ (109.23HN.002), August $11$ $2022$ (109.23HN.003), June $23$ $2023$ (111.254J.002), August $11$ $2023$ (111.254J.003), and August $06$ $2024$ (113.26GE.003) is Nortmann, while PIs of the nights of March $12$ $2022$ (108.22CH.001) and May $27$ $2022$ (109.232F.004) are Nagel and Diamond-Lowe, respectively.}
\end{table*}

\section{Methods}
\label{sec:methods}

Before performing any further operations in the data, we took into account the differences in the pixel-wavelength solution between the nodding A and B position spectra by interpolating the pixel-wavelength solution from nodding position B to position A. In addition, we removed the edges of each spectral order (the first and last $100$ pixels), as they show increased systematics. With all the spectra in a common wavelength solution (that of A spectra), we proceeded with the usual steps for our purpose with high-resolution spectra.

\subsection{Normalisation, outliers, and masking}
\label{subsec:methods_norm}

Ground-based observations of exo-atmospheres suffer from an inherent disadvantage stemming from the varying conditions of Earth's atmosphere (e.g., seeing and precipitable water vapour variability). Some spectra presented anomalous spectral behaviour, with unexpected peaks or valleys in the recorded counts, potentially caused by transient high-altitude clouds in Earth's atmosphere. Other spectra displayed a sharp decrease in mean signal-to-noise ratio compared to those taken immediately before and after. In these cases, we discarded the spectra from the analysis to avoid hindering the telluric correction (see discarded spectra listed in Table\,\ref{tab:observations}). In addition, variations during the observations induce a fluctuating baseline in the spectra (Fig.\,\ref{fig:steps}B). Thus, to provide a common continuum to all spectra, we performed an order-by-order normalisation by fitting a third-degree polynomial to the pseudo-continuum (Fig.\,\ref{fig:steps}C).

Next, following the approach used by \cite{2021ApJ...908L..17K}, we removed any possible contamination coming from cosmic rays by applying a $3\sigma$-clipping procedure where we flagged values that significantly deviate from the mean of the pixel's time series. We acknowledge that this procedure's threshold is conservative compared others (e.g., $5\sigma$-clipping). However, given the high S/N ratios of our datasets and the exposure times used, the distribution of the normalized flux at each pixel is expected to be well approximated by a Gaussian distribution. Moreover, previous studies have successfully applied this method using a comparable number of spectra per dataset \citep{2021ApJ...908L..17K, 2022A&A...666L...1S, 2022A&A...664A.121C, pelaez2025tighter}. Flagged values were corrected by interpolating over their nearest neighbours in wavelength. The most opaque windows of the Earth's atmospheric transmittance (telluric) were masked whenever $85$\% of the normalised flux was absorbed and also interpolated over their nearest neighbours. A safety window of $5$ pixels was added to each flagged pixel to further limit potential contamination of strong telluric-line wings. 

\subsection{Telluric and stellar correction}
\label{subsec:methods_telluric}

Even after masking the most opaque windows, telluric and stellar features within the spectra are several orders of magnitude stronger than the Doppler-shifted excess absorption from the planetary atmosphere. In order to disentangle the exo-atmospheric signal, we used {\tt SysRem} \citep{2005MNRAS.356.1466T, 2007ASPC..366..119M}, an iterative principal component analysis algorithm that accounts for unequal uncertainties for each data point. {\tt SysRem} has been successfully applied in similar studies in the past to reveal signatures in the atmosphere of several exoplanets \citep{2013A&A...554A..82D, 2013MNRAS.436L..35B, 2017AJ....153..138B, 2019A&A...630A..53S, 2020ApJ...898L..31N, 2022A&A...668A..53C, 2024A&A...692A...8M, 2025A&A...693A.213N}.

Contamination at different levels affect each spectral order depending on the spectral shape of the telluric transmittance. If a single global number of \texttt{SysRem} passes is adopted, there is a risk of over- or under-correcting for telluric features, which could either erase the planetary signal or leave residual contamination that hinders detection. This results in the number of required {\tt SysRem} passes being order-dependent and a criterion needs to be set to determine it. This is a difficult step that can lead to unintended biases and even to spurious planet-like signals, depending on the employed criterion \citep[see, e.g.,][]{2019MNRAS.482.4422C, 2023MNRAS.522..661C}. We therefore determined the appropriate number of \texttt{SysRem} passes individually for each spectral order by following a similar approach to \cite{2020AJ....160...93H, 2022AJ....163..248H, 2021AJ....161..209D, 2024AJ....168..106R} and \citet{2025MNRAS.538.3263P}. We studied the variance of the residual spectral matrices after applying {\tt SysRem} and halted the algorithm when the standard-deviation difference between two consecutive passes ($\Delta\sigma$) was below a given threshold ($1\%$) and started to plateau. In practice, this method helps recognise when $\Delta\sigma$ plateaus for each spectral order, indicating that the major spectral variations have been removed and no further passes are required. This should also preserve signals close to the noise level such as the exo-atmospheric absorption, while also being a model-independent approach, in contrast to optimisation-based on injection-recovery techniques. The $\Delta\sigma$ metric is thus defined as:

\begin{equation}
   \Delta \sigma=\frac{{ }^{(i-1)} \sigma-{ }^{(i)} \sigma}{^{(i-1)} \sigma},
\end{equation}

\noindent where $^{(i-1)} \sigma$ and $^{(i)} \sigma$ are the standard deviations of the residuals in a spectral order before and after the $i$-th pass of {\tt SysRem} respectively \citep{parker2025limits}.

\begin{figure}
    \centering
    \includegraphics[width=\columnwidth]{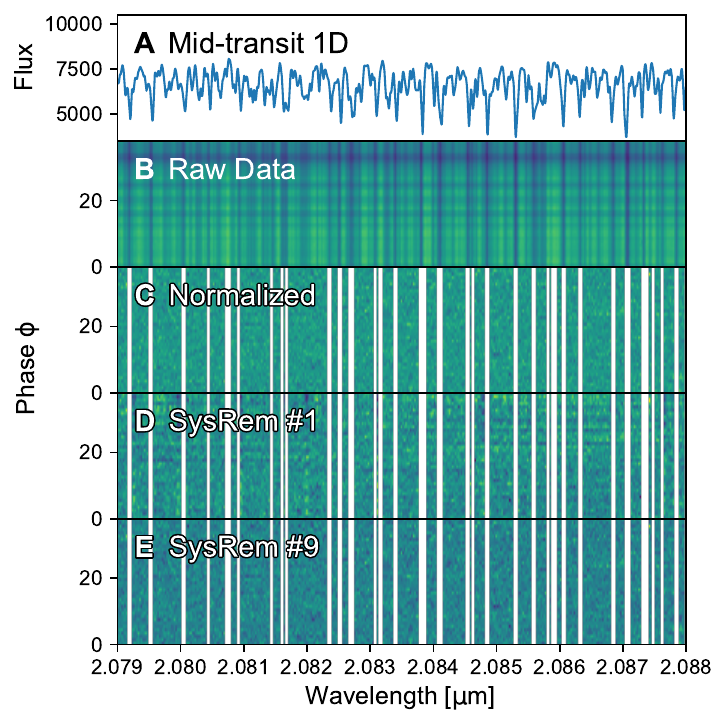}
    \caption{Steps of the data preparation applied to a representative spectral region of the dataset obtained on the night of March $31$, $2022$. The vertical axis in panel A represents flux in arbitrary units, while in all other panels it represents time, expressed as the planet's orbital phase. Panel A: Original spectrum observed at mid-transit, where telluric absorption lines from H$_2$O can be identified. Panel B: Original spectral matrix extracted using \texttt{CR2RES}, where major S/N differences between spectra (horizontal stripes) and prominent telluric H$_2$O absorption (vertical stripes) can be identified. Panel C: Normalised and masked spectra, where opaque telluric windows are excluded (white stripes). Panel D: Residual spectra after one {\tt SysRem} pass, where telluric residuals can be observed. Panel E: Residual spectra after nine {\tt SysRem} passes, where most of the telluric contribution has been removed. Nodding position effects were corrected in all panels by interpolating the pixel-wavelength solution from nodding position B to position A.}
    \label{fig:steps}
\end{figure}

In the defined metric, a plateau is typically reached between five to seven {\tt SysRem} passes for CRIRES$^+$ datasets, depending on the spectral order, as illustrated in Fig.\,\ref{fig:ds_per_order}. This indicates the number of \texttt{SysRem} passes adopted for each spectral order. During the first {\tt SysRem} pass, telluric and stellar features still remain (panel D of Fig.\,\ref{fig:steps}), suggesting the need for further cleaning. 
For intermediate to final passes, the exo-atmospheric signal, if present, is expected to be buried in the noise (Fig.\,\ref{fig:steps}E).

\subsection{Model atmospheres}
\label{subsec:methods_models}

We modelled the exo-atmospheric transmission spectra of GJ\,1214\,b using \texttt{petitRADTRANS}\footnote{\textcolor{magenta}{\texttt{https://petitradtrans.readthedocs.io}}} ({\tt pRT}), a Python-based package for calculating transmission and emission spectra of exoplanets through radiative transfer \citep{2019A&A...627A..67M}. We assumed H$_2$--He dominated atmospheres where Rayleigh scattering and collision-induced absorption are included \citep{2009A&A...493..671F, 2010A&A...523A..26N}. The same preparation steps performed on the real data were applied to the models to ensure that the same distortions are introduced.

We included in the \texttt{pRT} models the absorption of the most relevant spectroscopically active species in the covered NIR range, that is H$_2$O, CO$_2$, CO, NH$_3$, H$_2$S, using the line lists presented by \cite{2010JQSRT.111.2139R}, and CH$_4$ by \cite{2020ApJS..247...55H}. Specifically, in Sect.\,\ref{sec:results} we search for H$_2$O, CO$_2$, and CH$_4$. Model transmission spectra were generated on a $2$D-grid spanning the pressure level of a potential cloud deck ($p_{\rm c}$) and the metallicity of the atmosphere ($Z$). This latter step was facilitated by the algorithm \texttt{easyCHEM}\footnote{\textcolor{magenta}{\texttt{https://easychem.readthedocs.io}}} \citep{2025JOSS...10.7712L}, which provides with the mass fractions of the atmospheric compounds and the mean molecular weight (thus, the atmospheric scale height) as a function of the metallicity, for an isothermal P-T profile at the equilibrium temperature of the planet (see Table\,\ref{tab:planets_params}). Throughout this work, we fixed the atmospheric carbon-to-oxygen ratio to a solar value of $0.55$ \citep{2009ARA&A..47..481A}, which agrees well (within uncertainties) with the JWST-derived values reported in \citet{ohno2025possible}. Our $p_{\rm c}$ grid spanned from $0.1$\,mbar to $1$\,bar, while the $Z$ grid covered from $1\times$\,solar to $1000\times$\,solar metallicity. Grid values were evenly spaced on a logarithmic scale, resulting in a total of $100$ templates per molecule.
Fig.\,\ref{fig:model_templates} illustrates the variations of the transit depth in the models due to changes in both $p_{\rm c}$ and $Z$ in the wavelength range covered by the K2148 setting of CRIRES$^+$. For each $Z$\,--\,$p_{\rm c}$ pair, we compute two types of transmission models: (i) those including all spectroscopically active species in the K2148 setting of CRIRES$^+$, which will be injected in the data prior to preparation for detectability studies in Sect.\,\ref{sec:results}, and (ii) models including only a specific species' opacity, which will be used as the CCF template. We note that, in this latter case, we still consider all species from \texttt{easyCHEM} to assess the atmospheric composition and the scale height appropriately. Further discussion is provided in Sect.\,\ref{sec:results}.

\subsection{Cross-correlation technique}
\label{subsec:methods_cc}

\begin{figure}
    \centering
    \includegraphics[width=\columnwidth]{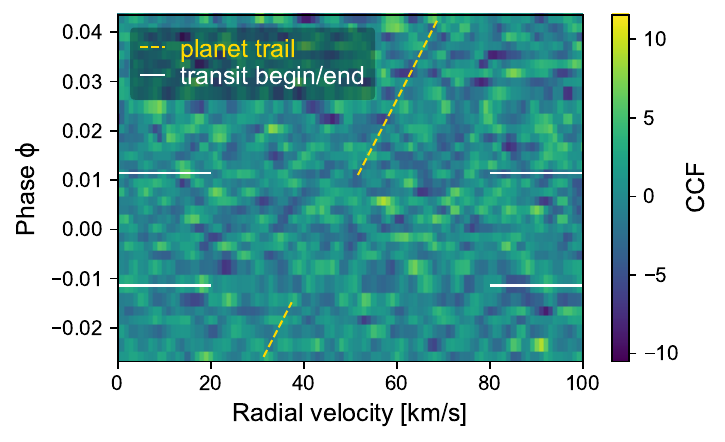}
    \caption{Cross-correlation analysis of potential H$_2$O signals in the transmission spectrum of GJ\,1214\,b observed with CRIRES$^+$ in the near-infrared. The method is illustrated for the night of March $31$, $2022$. We show the cross-correlation matrix in the Earth's rest frame as a function of the velocity Doppler shifts applied to the template (horizontal axis) and the planet's orbital phase (vertical axis). The results were obtained by using a template with  $10\times$\,solar metallicity and a $10$\,mbar cloud deck. All useful spectral orders were combined. White lines mark the transit start and end times, while the yellow lines indicate the expected exoplanet velocities with respect to Earth.}
    \label{fig:cc_matrix}
\end{figure}

Any potential atmospheric lines from GJ\,1214\,b would remain buried below the noise level in the residual spectra. By using the cross-correlation technique \citep{snellen2010orbital}, it is possible to co-add the contribution from potentially hundreds of these lines in the form of a cross-correlation function (CCF) peak \citep[e.g.,][]{2018arXiv180604617B, 2025ARA&A..63...83S}. We performed the cross-correlation between the residual matrices obtained in the previous steps $R_{ij}$ (where $i$ denotes each spectrum and $j$ the wavelength dimension) and the transmission models (templates) $m_j$ calculated using {\tt pRT}. In order to explore a wide range of velocities with respect to the Earth, we used linear interpolation to Doppler-shift the templates in a range of velocities $v$ from $-325$ to $+325$\,km\,s$^{-1}$ with respect to the Earth. We used $1$\,km\,s$^{-1}$ intervals, as set by the average velocity step size between the instrument's pixels. Given the CCF- and S/N-based metric used in this work, we do not expect any oversampling even in regions where the pixel velocity spacing becomes larger than the adopted step-size ($1$\,km\,s$^{-1}$) \citep{sanchez2025robustness}. CCFs were obtained individually for each spectrum, $i$, forming a cross-correlation matrix, $CCF(v,i)$, as follows:

\begin{equation}
\label{eq:ccf_vi}
    CCF(v,i) = \sum_j \sum_\lambda \frac{R_{i,\,j}(\lambda)\,m_j(\lambda,\,v)}{\hat \sigma_{i,\,j}^2 (\lambda)},
\end{equation}

\noindent where $\hat \sigma (\lambda)$ are the propagated uncertainties of the data and the summations loop over wavelength ($\lambda$) and spectral order $j$, so as to co-add the potential information contained in our full spectral coverage. The resulting total $CCF(v,i)$ in the Earth's rest frame is illustrated in Fig.\,\ref{fig:cc_matrix}, together with the trace where GJ\,1214\,b would be expected to appear during the transit (i.e., between the horizontal dash-dotted red lines) along the planetary RVs with respect to the Earth, $v_{\rm p}$. To describe the orbit of GJ\,1214\,b, we used the definition of $v_{\rm p}$ as follows:
\begin{equation}
    v_{\mathrm{p}}(\phi)=K_{\mathrm{p}} \cdot \sin2\pi\phi-v_{\text {bary }}+v_{\text {sys}},
    \label{eq:vp}
\end{equation}

\noindent where $\phi$ is the orbital phase, $v_{\rm bary}$ the barycentric velocity due to Earth's motion around the Solar System's barycentre, $v_{\rm sys}$ the systemic velocity of the star-planet system, and $K_{\rm p}$ the radial velocity semi-amplitude of the planet, defined as:
\begin{equation}
    K_{\mathrm{p}}=\frac{2 \pi a}{P_{\rm orb} \sqrt{1-e^2}} \sin(i).
    \label{eq:kp}
\end{equation}

\noindent where $a$ is the orbital parameters (the semi-major axis), $P_{\rm orb}$ the orbital period, $e$ the eccentricity, and $i$ the inclination. It is at this point that we visually inspected the cross correlation matrices in the Earth's rest-frame for all spectral orders and for each night separately (see Fig\,\ref{fig:cc_erf_orders}). We discarded from the analysis one spectral order with uncorrected systematics (see Table\,\ref{tab:observations}).

Next, we co-added all in-transit CCFs along the time axis (i.e., sum Eq.\,\ref{eq:ccf_vi} over $i$) to maximise any potential signal. To do this, we first Doppler-shifted $CCF(v,i)$ to the exoplanet's rest-frame by using Eq.\,\ref{eq:vp}, assuming $K_{\mathrm{p}}$ as unknown and using a range of values from $-300$ to $300$\,km\,s$^{-1}$ in steps of $1$\,km\,s$^{-1}$. Thereby, an absorption with a planetary origin should only appear around the expected $K_{\mathrm{p}}$ of GJ\,1214\,b (i.e., the exoplanet's rest frame) in the form of a CCF peak.

\begin{figure}
    \centering
    \includegraphics[width=\columnwidth]{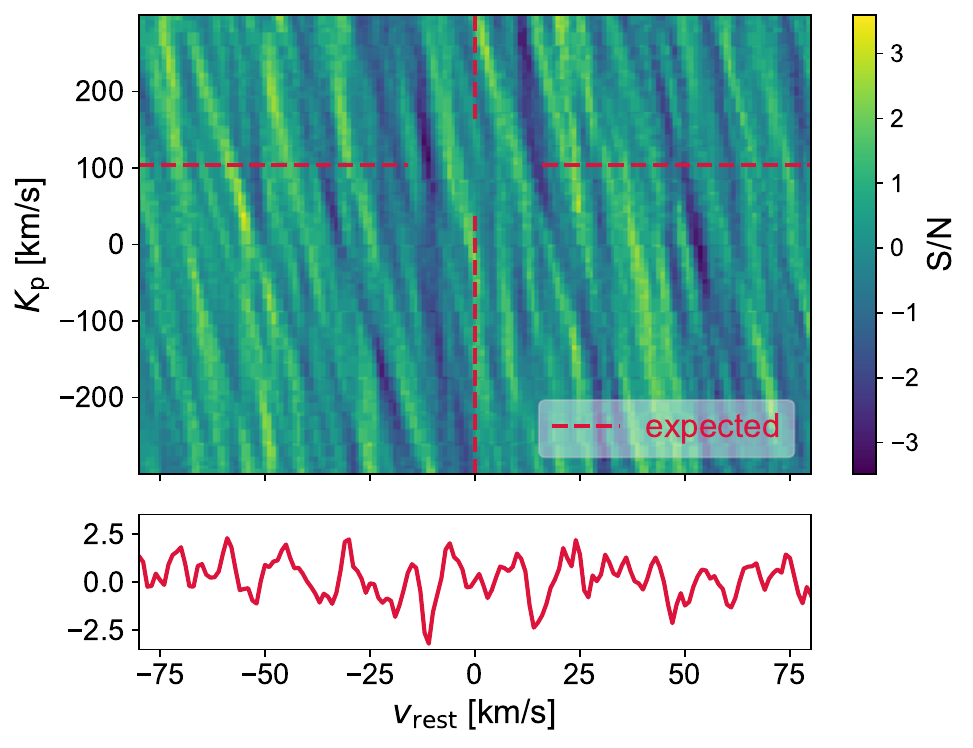}
    \caption{Significance analysis of potential H$_2$O signals, illustrated for the night of March $31$, $2022$. Top panel: signal-to-noise ratio map as a function of radial velocity in the exoplanet's rest frame ($v_{\rm rest}$, horizontal axis) and the projected orbital velocity semi-amplitude, $K_{\rm p}$ (vertical axis). Horizontal lines mark the expected $K_{\rm p}$, and vertical lines show the zero rest-frame velocity. Bottom panel: $1$D cross-correlation function at the expected $K_{\rm p}$ ($97.1$\,km\,s$^{-1}$) of GJ\,1214\,b.}
    \label{fig:snr_matrix}
\end{figure}

\subsection{Assessment of potential signals}
\label{subsec:methods_significances}

Following common signal significance evaluation approaches, we computed a S/N map as a function of the exoplanet rest-frame velocity and $K_{\rm p}$ (top panel of Fig.\,\ref{fig:snr_matrix}; \citealp{2017AJ....153..138B, 2018A&A...615A..16B, 2019A&A...621A..74A, 2019A&A...630A..53S, 2021ApJ...910L...9N, 2022A&A...668A..53C, 2024A&A...688A.206C}). For each $K_{\rm p}$, we divided each CCF point by the CCF's standard deviation, excluding a $\pm 20$\,km/s window around it. The division ensured that we avoided inflating the standard deviation by including potential CCF-peak wings. Potential signals are hence evaluated in a common $K_{\rm p}$--$\rm v_{rest}$ map, as shown in the top panel of Fig.\,\ref{fig:snr_matrix}. In this case, the map does not reveal any significant signals (e.g., at S/N$>$\,$4$) at the expected $K_{\rm p}$ (bottom panel of Fig.\,\ref{fig:snr_matrix}). All the observed patterns could be well explained by noise fluctuations or weak telluric residuals. This then leads to a non-detection of water vapour in GJ\,1214\,b when using this particular CRIRES$^+$ dataset observed on March $31$, $2022$. The same methods are applied subsequently to all transit datasets (see Fig.\,\ref{fig:snr_individual}).

To enhance the recovery of a potential planetary signal, we combined the information from each observed night by creating a merged CCF map. In this process, we sorted the frames from all datasets along the time axis according to their corresponding orbital phases following the approach of \cite{2022A&A...668A..53C, cont2025}. The respective $v_{\rm bary}$ of each spectrum was considered at the time of Doppler-shifting the CCF matrix from the Earth's rest-frame to the planet's. This procedure hence yielded a single CCF map, which incorporates all the data from the different nights and is used to compute the $K_{\rm p}$--$v_{\rm rest}$ and S/N maps in the same way as for a single-night dataset.

\begin{figure*}
    \centering
    \includegraphics[width=\textwidth]{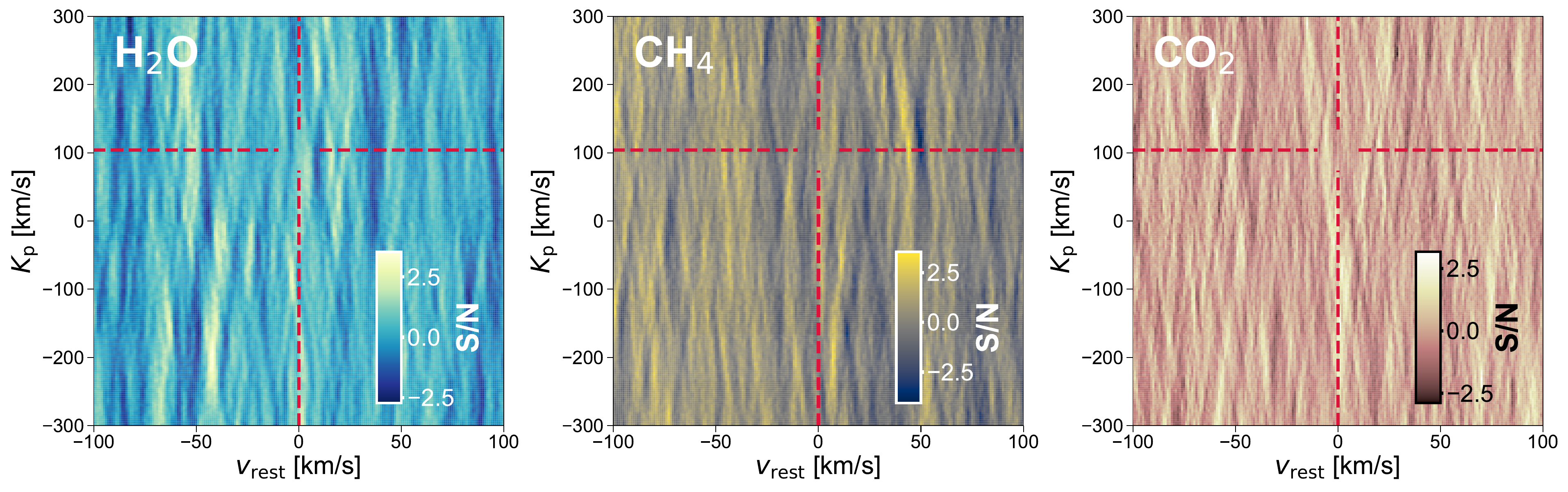}
    \caption{Cross-correlation maps in S/N units for potential atmospheric signals as a function of $v_{\rm rest}$ and $K_{\rm p}$ for the primary expected absorbing species in the atmosphere of GJ\,1214\,b. We explored, water vapour (left), methane (middle), and carbon dioxide (right). Horizontal and vertical red lines indicate the expected $K_{\rm p}$ and $v_{\rm rest}$. These example maps were derived using a $10\times$\,solar metallicity template with a cloud deck at $10$\,mbar. No molecular signals were detected in our cross-correlation analyses.}
    \label{fig:snr_matrices}
\end{figure*}

\section{Results and discussion}
\label{sec:results}

When performing direct cross correlation of the residual data after {\tt SysRem} with H$_2$O, CH$_4$, and CO$_2$ templates ($p_c$\,$=$\,$10$\,mbar, Z\,$=$\,$10\times$\,solar) for the eight transit observations of GJ\,1214\,b with CRIRES$^+$ we were not able to detect either water vapour, methane, or carbon dioxide in the atmosphere of this planet (see Fig.\,\ref{fig:snr_matrices}). That is, all signals in the map have significances around or below the S/N\,$=$\,$3$ level, hence being consistent with noise. This is in agreement with most of the previous observations of this exoplanet's atmosphere, which is historically reported to be cloudy, thus presenting muted spectral features. Unfortunately, this also means we were not able to confirm at high-resolution and using transit spectra any of the recently reported hints of molecular features from JWST data \citep{2023Natur.620...67K, 2024ApJ...974L..33S, 2025ApJ...979L...7O}.

\begin{figure}
    \centering
    \includegraphics[width=\columnwidth]{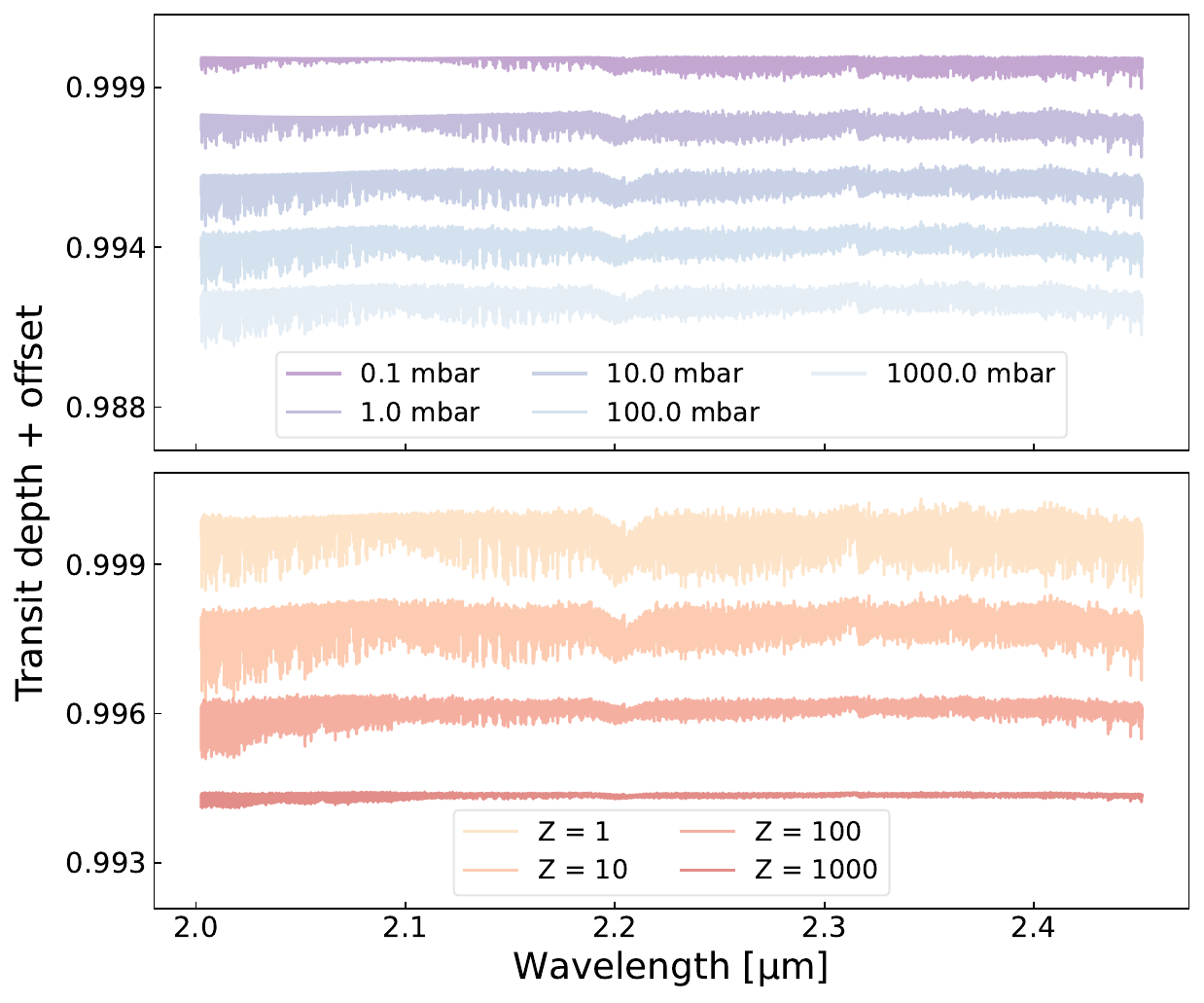}
    \caption{Transit depth of the main absorbers (H$_2$, He, H$_2$O, CO$_2$, CO, CH$_4$, NH$_3$, and H$_2$S) for GJ\,1214\,b in the analysed K-band range as a function of cloud deck pressure level (top panel, fixed metallicity of $10\times$\,solar) and atmospheric metallicity (bottom panel, clear atmosphere) as multiples of solar. For clarity, an offset of $0.002$ in transit depth was applied to the transmission models in both panels.}
    \label{fig:model_templates_complex}
\end{figure}

\begin{figure*}
    \centering
    \includegraphics[width=\textwidth]{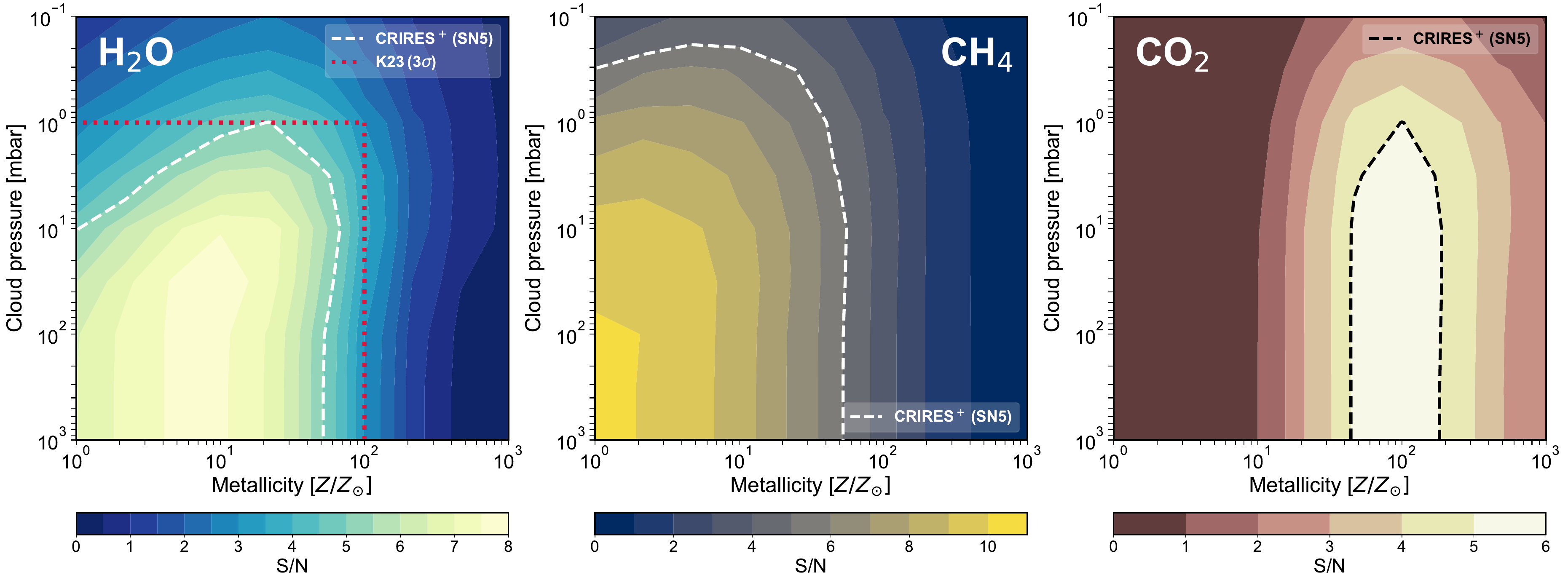}
    \caption{Detectability maps obtained from injection recovery tests, showing upper limits on the abundances of the main near-infrared strongest absorbers in the atmosphere of G\,1214\,b. Specifically, we investigate H$_2$O (left panel), CH$_4$ (middle panel), and CO$_2$ (right panel) across a range of metallicities and cloud deck altitudes. Dashed lines indicate the S/N\,$=$\,$5$ contours for H$_2$O, CH$_4$, and CO$_2$, as obtained from CRIRES$^+$ (white, eight transits combined). The lower S/N of CO$_2$ recoveries stems from its weaker and fewer lines in the K band, compared to the other infrared absorbers. Red dotted lines indicate the upper limits derived in \cite{2023Natur.620...67K} from day and nightside evidences of H$_2$O observed with JWST, in the best-fitting case of reflective clouds and high metallicity.}
    \label{fig:injections}
\end{figure*}

As it is a standard approach, the lack of molecular signatures in our K-band spectra motivated the exploration of the $Z$ and $p_c$ parameter spaces. Different methods have been used to construct the $Z$-$p_c$ grid to date: (1) a direct S/N metric where the grid is built using the pure S/N value \citep{grasser2024peering, pelaez2025tighter}; (2) an alternative $\Delta\sigma$ based analysis where the S/N value used in the grid comes from a likelihood framework (Log(L)) \citep{parker2025limits}; and (3) a Log(L) grid in which the quality of the fit of each tested model is compared against the best fitting model \citep{lafarga2023hot, dash2024constraints}. In this study, we have used method (1). Thus, using the previously generated set of models (see Sect.\,\ref{subsec:methods_models}), we performed injection-recovery tests to assess the detectability of different atmospheric properties in the planet with our combined observations (direct S/N metric). The injected model atmospheres hence include a full mean-molecular weight and mass fraction assessment by {\tt easyCHEM}, plus the opacity contributions from H$_2$, He, H$_2$O, CO$_2$, CO, CH$_4$, NH$_3$, and H$_2$S in the transmission spectrum calculation from {\tt petitRADTRANS} (Fig.\,\ref{fig:model_templates_complex}). As for the CCF template, it only includes one opacity source, that of the species to be recovered (see Sect\,\ref{sec:methods}). Using only the opacity of one molecular absorber (besides H$_2$) in the template reproduces the usual approach of CCF studies, where single-species templates are cross-correlated with exo-atmospheric data containing a wealth of spectral contributions, hence avoiding overly optimistic detectability estimations \citep[see discussion in][]{2025MNRAS.538.3263P}. The resulting detectability maps are shown in Fig.\,\ref{fig:injections} for the cases of H$_2$O, CH$_4$, and CO$_2$.

Methane explorations placed the strongest upper limits (S/N\,$=$\,$5$ contour), discarding atmospheres with $p_c$\,$\lesssim$\,$10^{-4}$\,bar for metallicities lower than $50\times$ solar. The parameter space discarded by H$_2$O analyses is fully contained in the CH$_4$-derived results. Both contours are in good agreement in their higher metallicity boundary ($50\times$ solar), but H$_2$O places a weaker restriction in the potential cloud deck location at all metallicities, and especially at solar values. This occurs because CH$_4$ lines in the K-band setting used in these observations contaminate the H$_2$O spectrum more strongly than the reverse, even though individual water vapour lines are intrinsically stronger.

Interestingly, CO$_2$ explorations yielded a surprisingly strong constraint (also a S/N\,$=$\,$5$ contour) discarding atmospheres for metallicities in the range $40$\,$<$\,$Z/Z_{\odot}$\,$<$\,$200$ and clouds at altitudes lower than the $1$\,mbar level. In this intermediate–to–high metallicity regime, and for the fixed C/O ratio of $0.55$ adopted in our chemical grid, the CO$_2$ abundance increases with the greater availability of carbon and oxygen atoms in the atmospheric reservoir. Consequently, the growth in CO$_2$ line opacity seems to outpace the loss of signal due to the reduced scale height at high metallicities. This balance, also suggested by \citet{parker2025limits}, opens a window for characterising more compressed atmospheres in the K-band for similar targets.

Overall, these results agree very well with the traditional observation-derived view of this exoplanet \citep[e.g.][]{bean2010ground, berta2012flat, kreidberg2014clouds}, which found its atmosphere to be covered by clouds that mute potential spectral features originating at altitudes below the cloud top deck. Our most-restrictive H$_2$O-derived upper limits agree very well with the best-fit scenario discussed in \cite{2023Natur.620...67K}, including high-altitude aerosols, a high-metallicity atmosphere, or both scenarios simultaneously.

It is also common to perform Bayesian retrievals to explore the parameter space we have discussed, since these represent powerful frameworks for assessing the properties of exoplanet atmospheres \citep{gibson2020detection, Brogi2019, blain2024formally, Lesjak2025Upper, pelaez2025tighter}. However, we did not consider that a Bayesian retrieval would provide additional meaningful constraints on the atmosphere of GJ\,1214\,b with our current datasets. 

In \citet{pelaez2025tighter}, a similar analysis combining high-resolution CARMENES and CRIRES$^+$ observations of GJ\,436\,b showed that Bayesian retrievals confirmed the non-detection of molecular species through cross-correlation analyses. In that study, the retrievals and injection–recovery experiments were found to be mutually consistent, with the retrieval posterior fully contained within the non-detectable region of parameter space as per CCF results. 
Comparable conclusions can be extracted from H$_2$O studies of KELT-11\,b and WASP-69\,b in \citet{Lesjak2025Upper}. 
Following the same rationale, we conclude that a Bayesian retrieval applied to our GJ\,1214\,b datasets would not yield additional constraints beyond injection–recovery tests.

Next-generation facilities like ANDES at the Extremely Large Telescope (ELT) are poised to revolutionise the field of exoplanet characterisation. To assess the capabilities of ANDES to probe the atmosphere of GJ\,1214\,b, we used {\tt EXoPLORE}, a comprehensive, end-to-end simulator of high-resolution observations, which is based on the techniques presented in \citet[][]{blain2024formally} and has been recently used for this aim in \citet{pelaez2025tighter}. {\tt EXoPLORE} is designed to produce realistic \emph{in-silico} time-series spectra, incorporating ANDES' wavelength coverage for the K-band (planned) and S/N based on ANDES exposure time calculator \citep{sanna2024andes} assuming exposure times of 30\,s, a resolving power of $\mathcal{R}$\,$=$\,$100,000$, and telluric evolution as a function of the airmass. It is important to note that the current version of the simulator does not yet incorporate additional sources of correlated noise or water vapour variability, which can greatly hinder the telluric correction. These remain active areas of development and, consequently, the results presented here should be considered as optimistic approaches to the real performance of ANDES. Using {\tt EXoPLORE}, we investigated the same grids in atmospheric metallicity and cloud-top pressure as in previous sections. For each combination, a complete simulated dataset was generated, incorporating four main spectral contributions:
\begin{itemize}
    \item The Earth’s atmospheric transmittance as a function of the airmass from ESO's Skycalc Tool \citep{noll2012atmospheric, jones2013advanced}.
    \item The exoplanet signal including H$_2$, He, H$_2$O, CO$_2$, CH$_4$, NH$_3$, and H$_2$S, Doppler shifted at each synthetic exposure following Eq.\,\ref{eq:vp} and scaled by the transit's light curve computed with the {\tt BATMAN} package \citep{kreidberg2015batman}, assuming a uniform stellar disk and the system parameters of GJ\,1214\,b listed in Table\,\ref{tab:planets_params} and from \citet{2024ApJ...963L..37M}.
    \item The host star, with a PHOENIX stellar template \citep{husser2013new} with T$_\mathrm{eff}$\,=\,$3000$\,K, $log(g)$\,=\,$5.0$, and [Fe/H]\,$=$\,$0.0$, consistent with the properties of GJ\,1214\footnote{The closest possible to the values T$_\mathrm{eff}$\,=\,$3026$\,K, $log(g)$\,=\,$5.026$, and [Fe/H]\,$=$\,$0.29$ from \citet{cloutier2021more}.}
    \item Random noise for each spectral point and exposure, generated using Python's functions {\tt np.random.default\_rng()} and {\tt rng.normal()}. Using the ETC, the standard deviation for the gaussian distribution was computed as $\sigma(\lambda)$\,$=$\,$1/\text{S/N}_{\rm ETC}(\lambda)$.  
\end{itemize}

Next, we analysed the simulated dataset following the techniques we presented in Sect.\,\ref{sec:methods}, and the signal recovered at the expected exoplanet $K_{\rm p}$ and $v_{\rm rest}$ was stored for the particular $(Z/Z_\odot,\,p_c)$ tuple. We note that, in consonance with our injection recovery tests (Fig.\,\ref{fig:injections}), the CCF templates were computed using the {\tt easyCHEM}-derived mean-molecular weight, including all equilibrium species, but using only single-species' opacity in {\tt petitRADTRANS}. It is instructive to place these simulations in the context of the existing VLT/CRIRES$^+$ observations. The number of detected photons scales as $N_\gamma$\,$\propto$\,$D^2\,T_{\rm exp}$, where $D$ is the telescope diameter and $T_{\rm exp}$ the exposure time. Using the exposure times adopted here (30\,s for the ELT and 240\,s for the VLT), the photon ratio per exposure (i.e., the collective power comparison) can be roughly estimated as
\begin{equation}
\frac{N_{\gamma,{\rm ELT}}}{N_{\gamma,{\rm VLT}}}
= \frac{D_{\rm ELT}^2\,T_{\rm exp,ELT}}{D_{\rm VLT}^2\,T_{\rm exp,VLT}}
\simeq \left(\frac{39}{8.2}\right)^2 \times \frac{30}{240}
\simeq 2.8.
\end{equation}
Thus, a single 30\,s ANDES exposure collects nearly three times more photons than a 240\,s CRIRES$^+$ frame. When comparing whole transits, the scaling simplifies if we assume similar overheads: ANDES observes one transit with exposure times eight times shorter, while we combined eight CRIRES$^+$ transit datasets that have, except for one night (2022 May 27, Table\,\ref{tab:observations}), exposures eight times longer. These two factors cancel each other, and the total photon ratio for an entire transit remains the same factor of $\sim$\,$2.8$.
In the photon-limited regime, the signal-to-noise ratio scales as $\sqrt{N_\gamma}$, so a single ANDES transit is expected to achieve a higher S/N per exposure by a factor of
\begin{equation}
\frac{\mathrm{S/N}_{\rm ELT}}{\mathrm{S/N}_{8\times{\rm VLT}}}
\simeq \sqrt{2.8} \simeq 1.7.
\end{equation}
That is, roughly a $70$\% improvement in line-contrast S/N compared to the full eight-transit CRIRES$^+$ data set. 
This estimate assumes comparable throughputs for ANDES and CRIRES$^+$ and therefore must be understood as a conservative, photon-limited comparison. It is important to caution that the injection–recovery analysis used for CRIRES$^+$ data differs in important ways from the idealised ANDES simulations presented here. In particular, the injected planetary signal to obtain the results of Fig.\,\ref{fig:injections} is not scaled by the transit light curve during ingress and egress, which artificially strengthens the recovered signal. Conversely, the VLT data contain real telluric and instrumental residuals that are not fully captured in our simulated spectra, although we did not observe significant telluric residuals in the included spectral orders across nights. These differences should be taken into account when comparing \textit{in-silico} against empirical results. \\

\begin{figure*}
\centering
\includegraphics[width=0.33\textwidth]{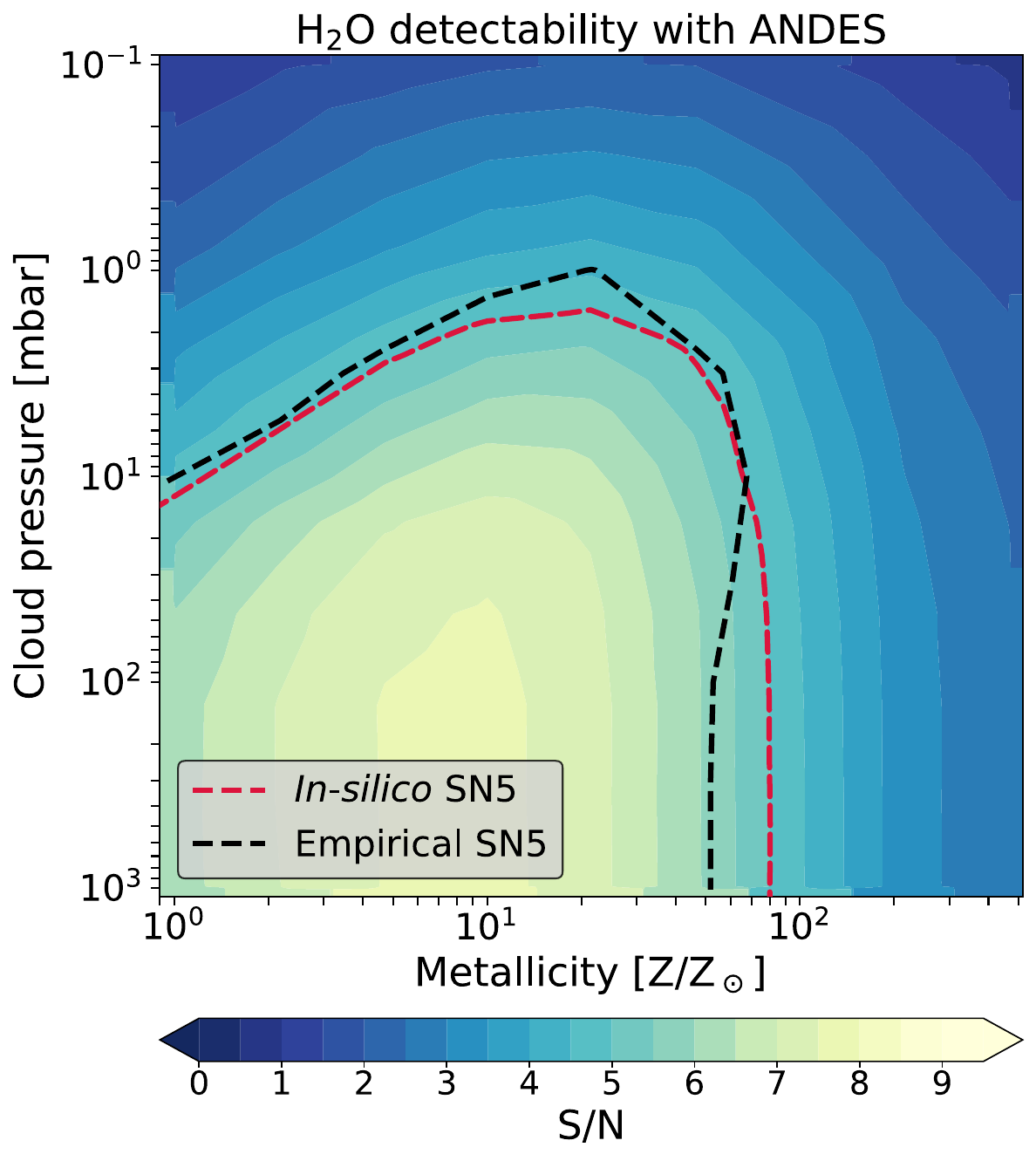}\,\includegraphics[width=0.33\textwidth]{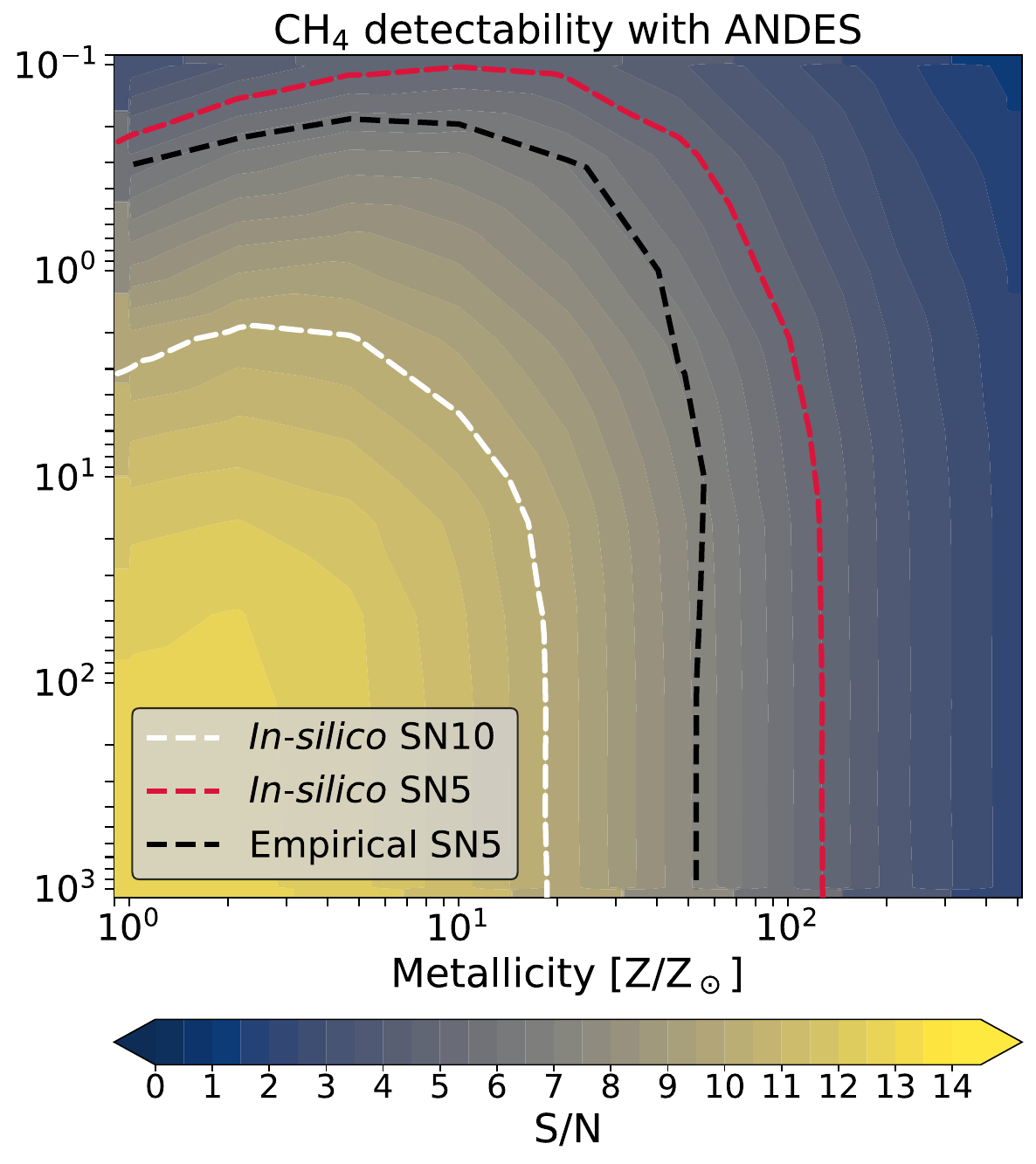}\,\includegraphics[width=0.33\textwidth]{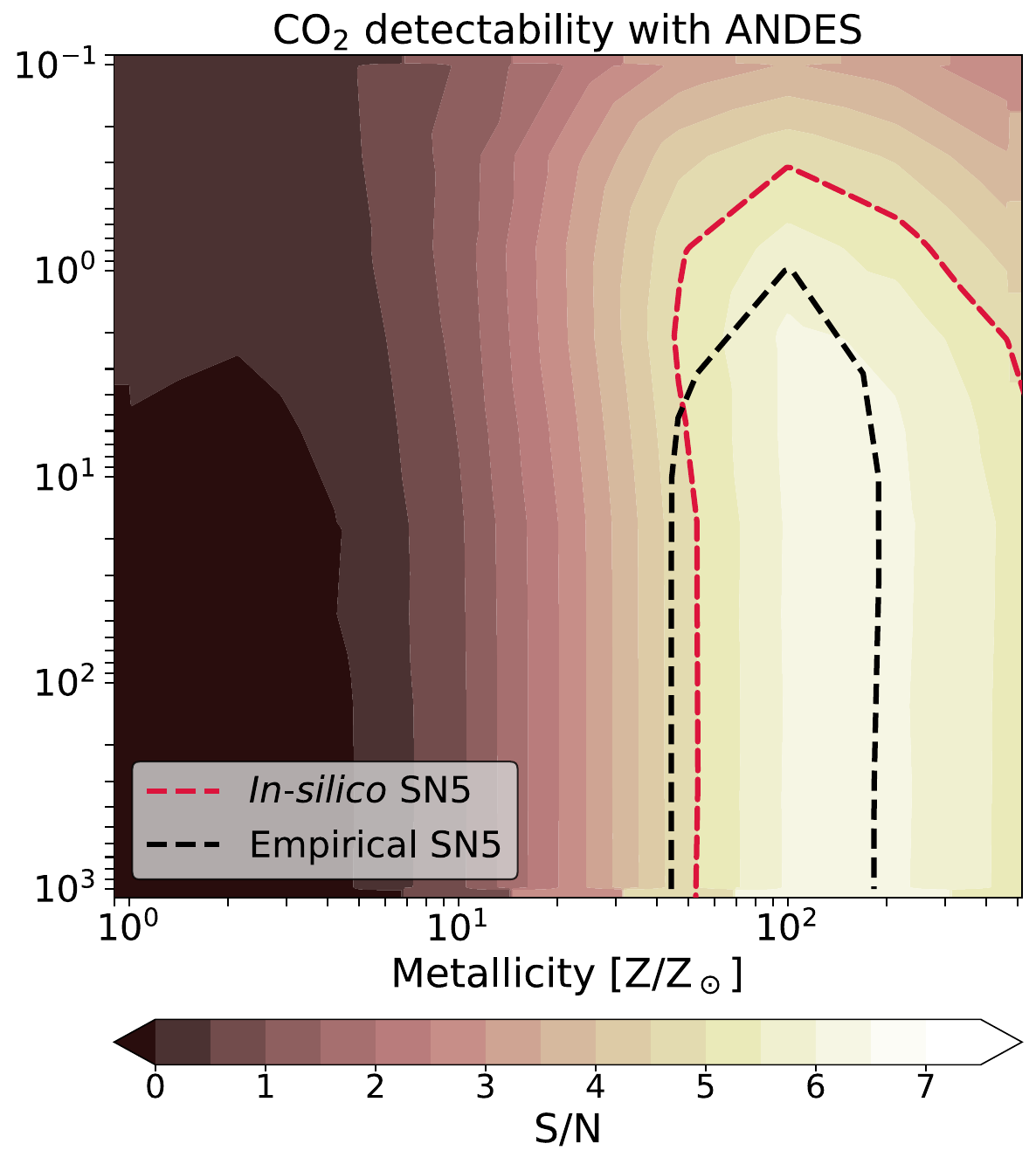}
\caption{Simulated detectability maps of GJ\,1214\,b computed with {\tt EXoPLORE}, assuming a single-transit observation with ANDES on the upcoming Extremely Large Telescope. We compute the case of H$_2$O (left), CH$_4$ (middle), and CO$_2$ (right). The white (only for CH$_4$) and red contours indicate recovered S/N levels of $10$ and $5$ in ANDES \textit{in-silico} data, respectively. For comparison purposes, we also include the  S/N\,$=$\,$5$ curves (black) obtained from the empirical data analysed in this work (Fig.\,\ref{fig:injections}).}
\label{fig:ANDES}
\end{figure*}

The predicted detectability maps obtained with ANDES, for a single transit observation of GJ\,1214\,b and searching for water vapour, methane, and carbon dioxide, are shown in Fig.\,\ref{fig:ANDES}.
A natural way to interpret the ANDES simulations is to compare them directly with the constraints obtained from the CRIRES$^+$ injection–recovery tests from Fig.\,\ref{fig:injections}. In all cases, the in-silico signal-to-noise maps for a single ANDES transit reproduce the overall structure of the empirically derived detectability contours, while extending the excluded region of parameter space in a way that reflects the underlying line strengths and chemical behaviour of each molecule.

For H$_2$O, the empirical and in-silico S/N\,$=$\,$5$ contours (SN5) are remarkably similar across the explored $Z$\,--\,$p_c$ plane, with only a modest shift towards slightly higher metallicities and somewhat deeper cloud decks at the high-$Z$ end. This is consistent with the H$_2$O not dominating the opacity in the K band at GJ\,1214\,b's equilibrium temperature. That is, H$_2$O lines are strong, but they are partially blended with CH$_4$ and other species and the significant overlapping with these other near-infrared absorbers shifts the detectability of H$_2$O from a photon-limited to a signal-limited regime, and hence its measurement does not benefit nuch from the increased collecting area of the ELT. In other words, the small difference between the empirical and the \textit{in-silico} contours indicates that we are already approaching a regime where H$_2$O detectability is limited by intrinsic atmospheric effects, namely the mixture of overlapping opacities and the reduced scale height at high metallicity, which together weaken the H$_2$O lines to undetectable levels. This contrasts with the photon-limited regime, where detectability could still be improved by collecting more photons.

Methane inferences behave more in line with a priori expectations from its line density in this wavelength range. In the CH$_4$ detectability map, the in-silico SN5 contour for a single ANDES transit excludes a significantly larger region of parameter space than the empirical CRIRES$^+$ one, particularly at high metallicities and for cloud decks deeper than $0.1$\,mbar. A S/N\,$=$\,$14$ threshold is reached \textit{in-silico}, compared to the maximum S/N\,$\approx$\,$10$ of the empirical case, which is consistent with the expectation that ANDES should deliver an overall (photon-limited) gain of order $\sim$\,$1.7$ in line-contrast S/N relative to the eight-transit CRIRES$^+$ set. 
This overall reflects the fact that CH$_4$ provides the densest and intrinsically strongest set of lines for CCF in the K-band for a cool, metal-rich atmosphere like that expected for GJ\,1214\,b.  

The behaviour of CO$_2$ further emphasises the role of chemistry in shaping the detectability we observed in Fig.\,\ref{fig:injections}. In the CO$_2$ maps, the empirical SN5 contour from CRIRES$^+$ data limits the parameter space to a region of intermediate metallicities around $100\times$\,solar and relatively low cloud-top pressures of $p_c$\,$<$\,$1$\,mbar. The \textit{in-silico} ANDES SN5 contour presents a similar morphology, modestly pushing upwards the $p_c$ boundary for $Z/Z_{\odot}$\,$=$\,$100$, while significantly expanding towards a broader range of metallicities, beyond $1000\times$\,solar for lower-altitude clouds. As discussed for empirical injection-recovery tests, this suggests that the steep increase in CO$_2$ abundance with metallicity translates into a line opacity in the K-band that rises strongly enough to compensate the high-$Z$ reduction in the features' amplitude. In that situation, the CO$_2$ signal would remain effectively photon-limited over a wider metallicity range and hence, increasing the collecting area directly impacts the detectability in compressed atmospheres with relatively deep clouds. The ANDES predictions therefore strengthen, rather than qualitatively change, the picture already suggested by the CRIRES$^+$ analysis, deeming CO$_2$ as an especially powerful tracer of high-$Z$ regimes in GJ\,1214\,b, and similar sub-Neptunes, within the K-band.

\section{Conclusions}
\label{sec:conclusions}

In this work, we have conducted the most comprehensive high-resolution spectroscopic study to date of the sub-Neptune exoplanet GJ\,1214\,b using eight transit observations obtained with CRIRES$^+$. By applying a rigorous telluric and stellar correction followed by cross-correlation analyses, we searched for atmospheric signatures of H$_2$O, CH$_4$, and CO$_2$ in the planet’s transmission spectrum across the K band. This study is motivated by the recent hints of these molecules from JWST Mid-Infrared Instrument (MIRI) observations in phase curves \citep{2023Natur.620...67K}, transmission spectroscopy \citep{2024ApJ...974L..33S}, and by combining HST, JWST/NIRSpec, and JWST/MIRI data \citep{2025ApJ...979L...7O}.

Our high-resolution analysis of eight CRIRES$^+$ transits combined of GJ\,1214\,b yields no significant detections of H$_2$O, CH$_4$, or CO$_2$, with all cross-correlation signals consistent with noise. Injection–recovery tests nonetheless place meaningful upper limits, ruling out a broad range of low-altitude cloud, low-metallicity scenarios and favouring atmospheres with high-altitude aerosols, high metallicity, or both. CH$_4$ provides the tightest constraints toward lower metallicities ($Z/Z_{\odot}$\,$<$\,$50$; $p_c$\,$>$\,$0.2$\,mbar), while CO$_2$ probes an intermediate-metallicity window at $Z/Z_{\odot}$\,$\approx$\,$100$, where its increased abundances seems to compensate the reduced scale heights arising from the higher metallicity values \citep[see also discussion in][]{parker2025limits}. These results are fully consistent with the aerosol-rich, high-metallicity atmospheric picture derived from previous studies and from recent JWST inferences for this exoplanet.

GJ\,1214\,b hence remains an archetype of an atmosphere with muted spectral features. Our results highlight the challenges of high-resolution spectroscopy for probing cloudy sub-Neptune atmospheres. Despite their much higher resolving power compared to JWST, current ground-based spectrographs still lack the sensitivity needed to detect the weak line cores that may emerge above the clouds in transmission in GJ\,1214\,b. The combination of multiple epochs improves sensitivity, but it is also hindered by the difficulties of mitigating telluric systematics in several observations. The ANDES simulations we have presented indicate that the ELT will mark an important revolution by reaching, in one transit observation, significantly better constraints than eight CRIRES$^+$ transits combined. For H$_2$O, the predicted detectability in GJ\,1214\,b and similar planets is only moderately improved relative to CRIRES$^+$, consistent with H$_2$O absorption in the K-band being partially blended with other species. CH$_4$, whose dense line forest dominates the K2148 wavelength setting we analysed, shows a clearer benefit from the increased collecting area, as ANDES recovers a substantially larger region of parameter space at CCF S/N thresholds of $5$\,--\,$10$. An outstanding improvement is also obtained for CO$_2$, whose abundance and line opacity increase steeply with metallicity. For this molecule, the ELT expands the excluded region well beyond what is accessible to eight combined CRIRES$^+$ transits. Taken together, these results show that while H$_2$O remains challenging in a compressed, cloudy atmosphere at moderate T$_{\rm eq}$ values around $600$\,K, CH$_4$ and especially CO$_2$ should become significantly more accessible with a single ANDES transit, opening a path to constraining high-metallicity scenarios that remain out of reach with current facilities.

\begin{acknowledgements}
      We thank Sophia Vaughan for helpful comments on this manuscript. IAA authors acknowledge financial support from the Severo Ochoa CEX2021-001131-S and PID2022-141216NB-I00 grants of MICIU/AEI/ 10.13039/501100011033. Part of this work was supported by \emph{ESO}, project
      number Ts~17/2--1.
      We acknowledge financial support from the Agencia Estatal de Investigaci\'on of the Ministerio de Ciencia e Innovaci\'on MCIN/AEI/10.13039/501100011033 and the ERDF “A way of making Europe” through project PID2021-125627OB-C32, and from the Centre of Excellence “Severo Ochoa” award to the Instituto de Astrofisica de Canarias.
\end{acknowledgements}

\bibliographystyle{aa}
\bibliography{biblio}

\onecolumn

\begin{appendix}
\section{Additional plots}
\label{app:additional_plots}

\begin{figure}[h!]
    \centering
    \includegraphics[width=0.8\textwidth]{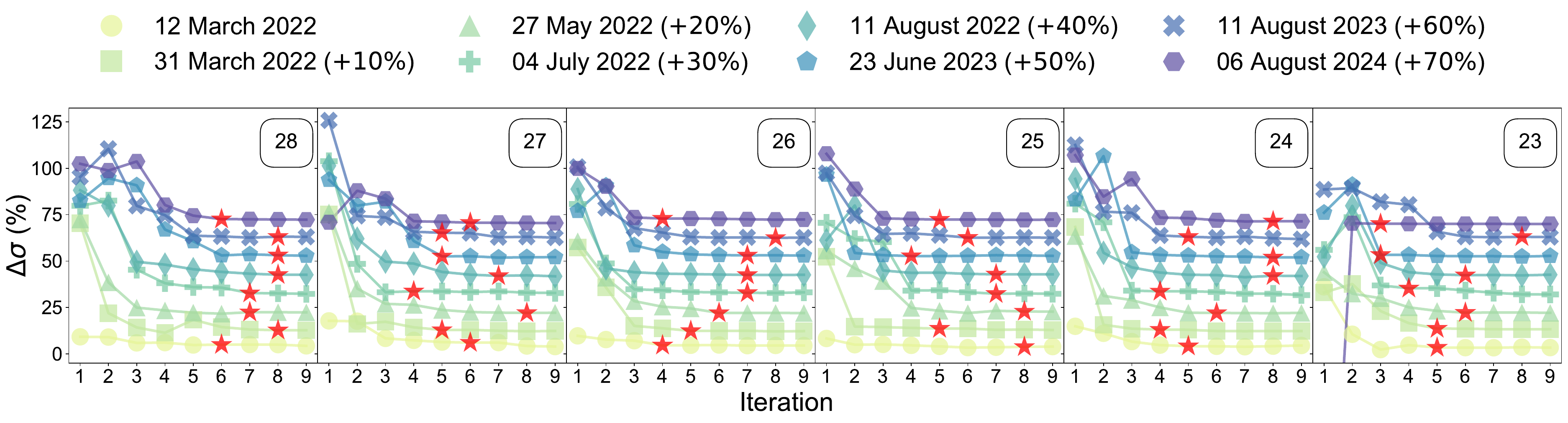}
    \caption{Illustration of the selection criterion used for order-wise SysRem optimisation for the CRIRES$^+$ datasets. Here, the values of $\Delta\sigma$ (vertical axis, where $\sigma$ is the standard deviation of the residual matrix; see text) are plotted as a function of the SysRem pass (horizontal axis) across all spectral orders for observations from all six nights. Red stars mark the SysRem pass at which a plateau is reached, indicating the point at which the algorithm is halted. Spectral order number is indicated in the upper right corner of each cell. For clarity, an offset was added to the per-night curves, as indicated in the legend.}
    \label{fig:ds_per_order}
\end{figure}

\begin{figure}[h!]
    \centering   \includegraphics[width=\textwidth]{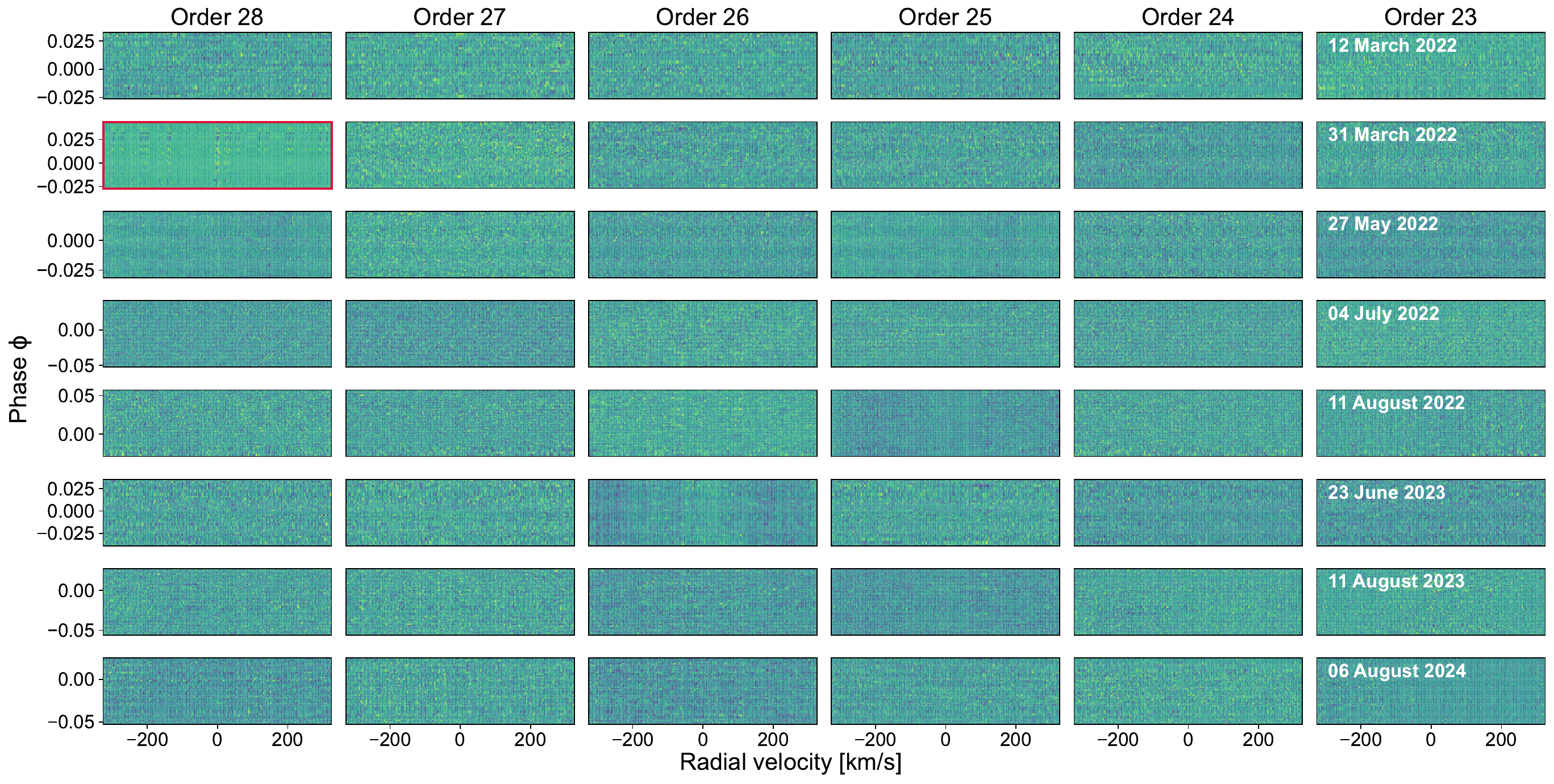}
    \caption{Cross-correlation matrices in the Earth's rest frame for potential H$_2$O signals in GJ\,1214\,b as a function of the velocity Doppler shifts applied to the template (horizontal axis, full range explored) and the planet's orbital phase (vertical axis). Results are shown per spectral order (columns) and for each night (rows). The H$_2$O template employed presented $10\times$\,solar metallicity and a $10$\,mbar cloud deck. The red box marks an excluded spectral order due to uncorrected systematics.}
    \label{fig:cc_erf_orders}
\end{figure}

\begin{figure}
    \centering
    \includegraphics[width=0.6\columnwidth]{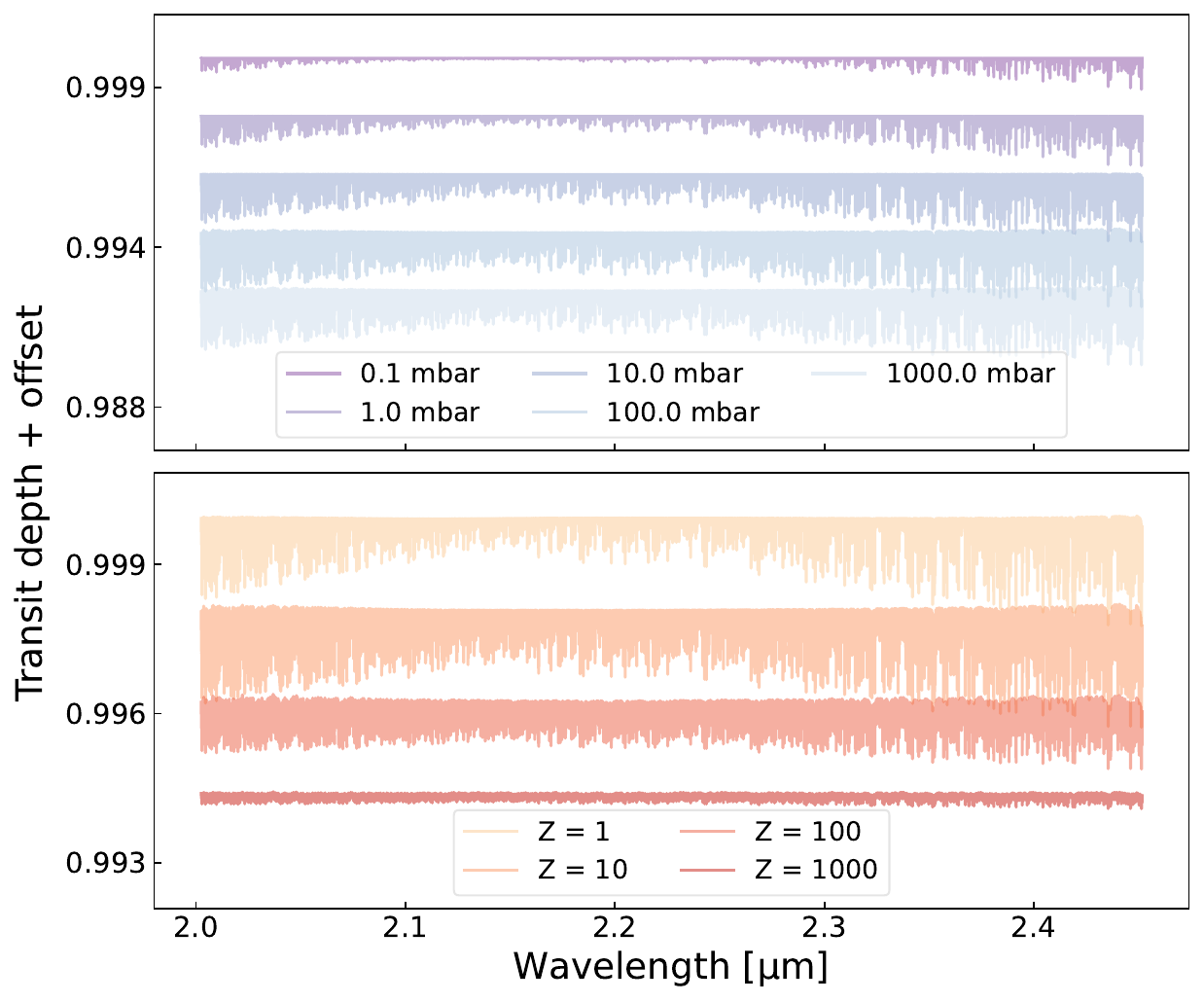}
    \caption{Transit depth of H$_2$O for GJ\,1214\,b in the analysed K-band range as a function of cloud deck pressure level (top panel, fixed metallicity of $10\times$\,solar) and atmospheric metallicity (bottom panel, clear atmosphere) as multiples of solar. For clarity, an offset of $0.002$ in transit depth was applied to the transmission models in both panels.
    }
    \label{fig:model_templates}
\end{figure}

\begin{figure}[h!]
    \centering   \includegraphics[width=\textwidth]{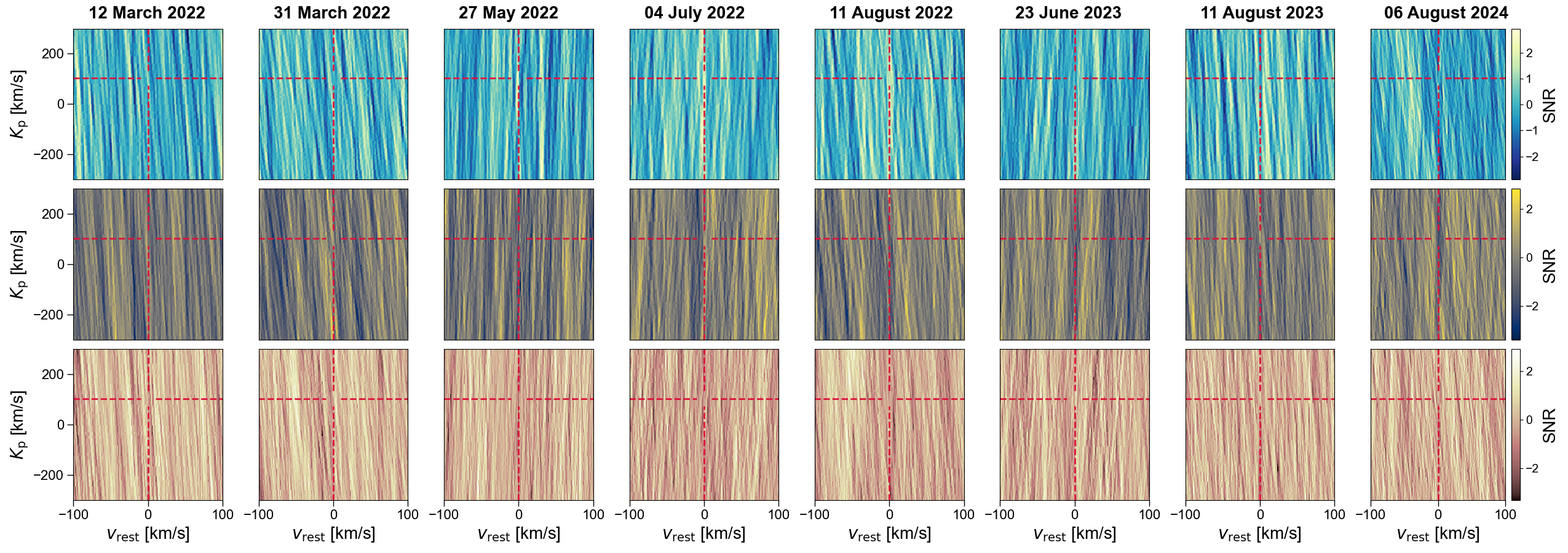}
    \caption{Same as Fig.\,\ref{fig:snr_matrix}, for each individual transit analysed and for the molecules studied. First row corresponds to water vapour, second row to methane, and third to carbon dioxide.}
    \label{fig:snr_individual}
\end{figure}

\end{appendix}

\end{document}